\documentclass[mathleft
]{an}
\usepackage{graphicx}
\usepackage{times}
\usepackage{natbib}
\bibpunct{(}{)}{;}{a}{}{,}
\begin{document}

\Pagespan{789}{}
\Yearpublication{}%
\Yearsubmission{}%
\Month{}%
\Volume{}%
\Issue{}%

\title{Complex asteroseismology of the hybrid B-type pulsator $\gamma$ Pegasi:\\ a test of stellar opacities}

\author{P. Walczak\thanks{\email{walczak@astro.uni.wroc.p}}
\and  J. Daszy\'nska-Daszkiewicz\thanks{\email{daszynska@astro.uni.wroc.p}\newline}}
\titlerunning{Complex asteroseismology of $\gamma$ Pegasi}
\authorrunning{J. Daszy\'nska-Daszkiewicz \& P. Walczak}
\institute{Instytut Astronomiczny, Uniwersytet Wroc{\l}awski, ul. Kopernika 11, 51-622 Wroc{\l}aw, Poland}

\received{}
\accepted{}
\publonline{later}

\keywords{stars: early-type -– stars: oscillations –- stars: individual: $\gamma$ Pegasi –- atomic data: opacities}

\abstract{ Using the updated oscillation spectrum of $\gamma$ Pegasi, we construct a set of seismic models which
reproduce two pulsational frequencies corresponding to the $\ell=0$, p$_1$ and $\ell=1$, g$_1$ modes.
Then, we single out models which reproduce other well identified modes.
Finally, we extend our seismic modelling by a requirement of fitting also
values of the complex, nonadiabatic parameter $f$ associated to each mode frequency.
Such complex asteroseismology of the B-type pulsators provides a unique test of stellar metallicity
and opacities. In contrast to our previous studies, results for $\gamma$ Peg indicate
that both opacity tables, OPAL and OP, are equally preferred.}

\maketitle

\section{Introduction}

Oscillation spectra of the main sequence B-type pulsators are characterized by
a low number of frequency peaks and a lack of distinct regularities.
Despite of these facts, asteroseismology of these massive stars
provides valuable constraints on stellar physics and evolution.

Last years, the most attractive targets for asteroseismic studies have become
hybrid pulsators which exhibit low order p/g modes as well as
high-order g-modes. This is because these two types of modes sound various
parts of the stellar interior. In the case of the early B-type stars,
modes typical for the $\beta$ Cep and SPB type occur simultaneously.
The analysis of extensive observations, especially from multisite campaigns,
allowed to discover several pulsating variables of the $\beta$ Cep/SPB type.

The first detected hybrid pulsator of the early B spectral type was $\nu$ Eridani
\citep{Hetal04,Aetal04,J05}.
The next example is 12 Lacertae \citep{Hetal06,Detal09}. Many excellent papers devoted to seismic modeling
of these stars have been published, but a limited space of this article does not allow to mention them.

$\gamma$ Pegasi is one more hybrid pulsator of early B spectral type.
Recent analysis of the space based observations from the MOST satellite \citep{Hetal09}
and ground based photometry and spectroscopy \citep{H09}
led to discovery of 14 pulsational frequencies with 8 of the $\beta$ Cep type and 6 of the SPB type.

Asteroseismic modeling, consisting in fitting theoretical and observational values of pulsational frequencies,
can be extended by adding another seismic tool
associated to each mode frequency. Such a tool is the amplitude of
the bolometric flux perturbation to the radial displacement, called
the $f$-parameter.
This seismic probe was introduced by \linebreak\citet{DDDP03,DDDP05}.
Theoretical values of $f$ are obtained
from stellar pulsation computations and their empirical
counterparts are derived from multicolour photometry.
The parallel fitting of pulsational frequencies and corresponding
values of the $f$-parameter was termed {\it complex asteroseimology}
by\linebreak \citet{DW09}. The hybrid pulsators are of particular interest
because the dependence of $f$ on pulsational frequency, $\nu$, and the mode degree, $\ell$, is completely different
for p modes and high-order g-modes.

Complex seismic modeling has been applied to the $\beta$ Cep star $\theta$ Oph \citep{DW09}
and to $\nu$ Eri \citep{DW10}. In the case of $\theta$ Oph, we got a strong preference for the OPAL data.
From the $\nu$ Eri analysis a contradictory result was obtained: the $\beta$ Cep-type modes indicate the OPAL opacities
whereas the SPB-type modes prefer the OP data.

In this paper we present complex seismic modeling of $\gamma$ Peg.
In Section 2, we give a short description of the star. Section 3 is devoted
to mode identification for all detected pulsational frequencies.
Results of our seismic modeling are presented in Sections 4.
Conclusions end the paper.

\section{$\gamma$ Pegasi}

$\gamma$ Peg (HD 886) is a bright star (V=2.83 mag) of the B2IV spectral type.
Its low amplitude variability was discovered almost a century ago
from the radial velocity observations \citep{B1911}.
These changes were confirmed by \citet{McN53}
who classified $\gamma$ Peg as the $\beta$ Cephei pulsating variable.

For dozens of years, $\gamma$ Peg has been thought to be a monoperiodic star,
pulsating in the radial mode with the period of about 0.15175 d \citep{J1970}.
Finally, \linebreak\citet{C06} reported discovery of three additional pulsational frequencies;
one in the $\beta$ Cep range and two in the SPB frequency domain.
\citet{C06} claimed also that this star is a spectroscopic binary with the orbital period of 370.5 d.
However, on the basis of space observations from the MOST mission and the ground based photometry
and spectro\-sco\-py, \citet{Hetal09} and \citet{H09} showed that $\gamma$ Peg is a single star
and the hypothetical orbital variations can be explained by the high-order g-mode pulsation.

\begin{figure}
\includegraphics[clip,width=83mm]{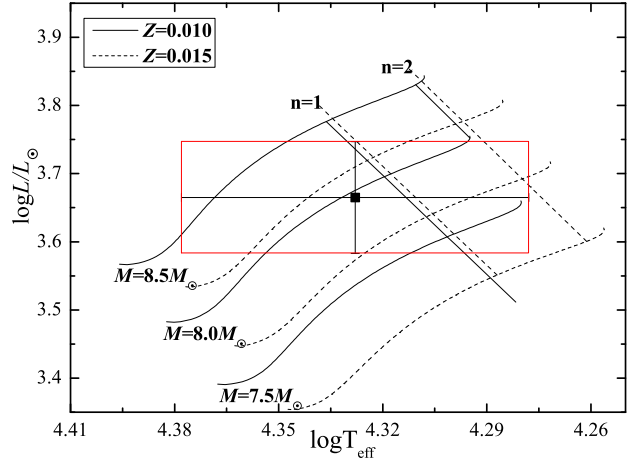}
\caption{The observational error box of $\gamma$ Peg in the HR diagram. The evolutionary tracks were computed
for two values of metallicity, $Z$. Lines of the constant
period (0.15175 d) for the fundamental ($n=1$) and first overtone ($n=2$) radial mode are also drawn.}
\label{HR}
\end{figure}

In Fig.\,\ref{HR}, we show a position of $\gamma$ Peg in the HR \linebreak diagram.
We included the whole range of the effective \linebreak temperature, $T_{\rm eff}$, available in the literature \linebreak
\citep[e.g.][]{DDN05, Metal06,HG08, H09} and the luminosity was calculated from the Hipparcos parallax and the bolometric correction from models of \citet{K04}.
The evolutionary tracks were computed with the War\-saw-New Jersey evolutionary code adop\-ting the OPAL opa\-cities \citep{OPAL}
and the newest heavy element mixture by \citet{AGSS09}, hereafter AGSS09.
Only evolution on main sequence is shown.
The star is a slow rotator with $V_{\rm{rot}}\sin{i}\approx 0$ as determined from the SiIII lines by \citet{Tetal06}.
The rotational splitting from the possible two components of the $\ell=1$ triplet gave $V_{\rm{rot}}\approx 3$ km/s
\citep{Hetal09}. Metallicity of $\gamma$ Peg is smaller than the \hbox{solar} value.
\citet{Metal06} determined \linebreak$Z=0.009\pm0.002$ from the optical spectra, whereas \linebreak
\citet{DDN05} obtained \linebreak $[m/H]=-0.04\pm0.08$ (equ\-ivalent to $Z=0.018\pm 0.003$)
from the IUE ultraviolet spectra.
The evolutionary tracks in Fig.\,1 were computed at the assumption of the equatorial rotational velocity
of 3 km/s, hydrogen abundance of $X=0.7$, two values of metallicity parameter $Z=0.010$
and 0.015 and no overshooting from a convective core, $\alpha_{\rm{ov}} = 0.0$.
Lines labeled as $n=1$ and $n=2$ will be discussed later on.

\section{Identification of the mode degree, $\ell$}

To determine the mode degree, $\ell$, of the observed frequencies of $\gamma$ Peg,
we made use of the light variations in the  Str\"omgren $uvy$ passbands \citep{H09}
and the radial velocity changes \citep{Hetal09}. To this aim we applied three approaches.

In the first case, we compared theoretical and observational values of the amplitude ratios and phase differences between the available passbands and relied on the theoretical values of the $f$-parameter. The second approach includes also the radial velocity variation. In the third method we used amplitudes and phases themselves and the $f$-parameter was determined from the observations together with the \linebreak mode degree, $\ell$ \newline\citep{DDDP03,DDDP05}. In the case of B-type pulsators, the last method demands the radial velocity measurements to get a unique identification of $\ell$.

In Table\,\ref{MI}, we give the most probable values of the mode degrees for the $\gamma$ Peg frequencies from these three approa\-ches.
The dominant mode, $\nu_1$, is certainly radial. The identification of $\nu_2$ is also unique; it is undoubtedly the $\ell=2$ mode.
In the case of $\nu_3$ we did not get a consistent result; it can be $\ell=1$ or $\ell=2$. The $\nu_4$, $\nu_6$, $\nu_8$ and $\nu_{12}$ modes have most probably the degree $\ell=2$. The $\nu_5$ and $\nu_{11}$ frequencies seem to be the dipole modes. Identification of the remaining frequencies are ambiguous, although in the case of $\nu_{14}$ only higher degrees are possible, with $\ell=6$ as most probable. It is interesting that $\nu_{14}$ is close to the first overtone radial mode in the $\gamma$ Peg models, but this identification is excluded
by our analysis.

\begin{table}[h]
\begin{center}
\caption{The most probable identification of $\ell$ for the pulsational frequencies
of $\gamma$ Peg from three approaches.}
\begin{tabular}{lccc}
\hline
frequency & phot.               & phot.+$V_{\rm{rad}}$&phot.+$V_{\rm{rad}}$\\
$[$c/d$]$& theoretical $f$ & theoretical $f$      &empirical $f$  \\ \hline
$\nu_{1}$=6.58974 &$\ell$=0    &$\ell$=0      &$\ell$=0    \\
$\nu_{2}$=0.63551 &$\ell$=2    &$\ell$=2      &$\ell$=2    \\
$\nu_{3}$=0.68241 &$\ell$=1    &$\ell$=2      &$\ell$=1,2  \\
$\nu_{4}$=0.73940 &$\ell$=2,1  &$\ell$=2      &$\ell$=1,2  \\
$\nu_{5}$=6.01616 &$\ell$=1,3  &$\ell$=1,2    &$\ell$=1,0,3\\
$\nu_{6}$=0.88550 &$\ell$=2    &$\ell$=2      &$\ell$=2,5  \\
$\nu_{7}$=6.9776  &$\ell$=?    &$\ell$=1,2,3,6&$\ell$=?    \\
$\nu_{8}$=0.91442 &$\ell$=2,3,4&$\ell$=1,2    &$\ell$=2,3,5\\
$\nu_{9}$=6.5150  &$\ell$$\ge$1&$\ell$=1,2,3,6&$\ell$=?    \\
$\nu_{10}$=8.1861 &$\ell$=?	   &$\ell$=1,2,3,6&$\ell$=?    \\
$\nu_{11}$=0.8352 &$\ell$=1,2  &$\ell$=1      &$\ell$=1,2  \\
$\nu_{12}$=6.0273 &$\ell$=?    &$\ell$=1,2    &$\ell$=2,5  \\
$\nu_{13}$=9.1092 &$\ell$=?	   &$\ell$=1,2,6  &$\ell$=?    \\
$\nu_{14}$=8.552  &$\ell$$\ge$4&$\ell$=6      &$\ell$=?    \\
\hline
\end{tabular}
\label{MI}
\end{center}
\end{table}

\begin{figure}
\begin{center}
\includegraphics[clip,width=83mm,height=105mm]{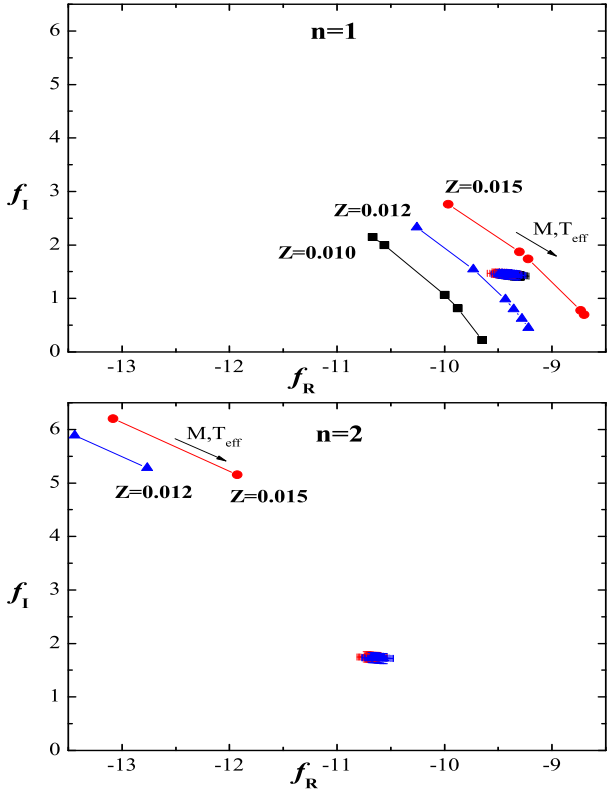}
\caption{Comparison of empirical and theoretical values of $f$ for the dominant frequency, $\nu_1$, on the complex plane.
The top and bottom panels correspond to the hypothesis of the fundamental and first overtone
radial mode, respectively.}
\label{fr-fi}
\end{center}
\end{figure}

In Fig.\,\ref{HR}, we plot lines of a constant period (0.15175 d) corresponding to $\nu_1$
for the fundamental and first overtone radial mode.
The discrimination of the radial order, $n$, can be done from a comparison of the empirical and theoretical values
of the $f$-parameter. This is shown in Fig.\,\ref{fr-fi}.
In the top panel, we plot the empirical and theoretical values of the $f$-parameter on the complex plane assuming
that the dominant mode is fundamental. The theoretical values of $f$ were calculated for three different
metallicities, $Z=0.010$, 0.012, and 0.015, overshooting parameter $\alpha_{\rm{ov}}=0.0$, hydrogen abundance $X=0.7$
and the OPAL data. In the bottom panel, we plot the same but we assumed that $\nu_1$ is the first overtone mode.
In both cases, we included models that are inside the observational error box of $\gamma$ Peg. As we can see,
an agreement between empirical and theoretical values of the $f$-parameter can be achieved only if $\nu_1$ is
the radial fundamental mode, $n=1$, and $Z$ is in the range (0.012, 0.015).

\section{Complex asteroseismology}

To compute pulsational models, we used the linear nonadiabatic code of \citet{D77}.
We started our seismic modeling by fitting two p-mode frequencies: $\nu_1=6.58974$ c/d and $\nu_5=6.01616$ c/d.
A survey of pulsational models showed that if $\nu_1$ is the radial fundamental mode, then $\nu_5$, identified
as $\ell=1$, can be only g$_1$. Here, we assumed that the azimuthal number of $\nu_5$ is $m=0$.
In all computations, we assumed the rotation velocity of $V_{\rm{rot}}=3$ km/s
and adopted the latest determination of the chemical composition AGSS09.
In the left panel of Fig.\,3, we present our seismic models fitting $\nu_1$ and $\nu_5$
on the $\alpha_{\rm{ov}}~ vs.~ Z$ plane for the hydrogen abundance of $X=0.7$ and the OPAL opacities.
\begin{figure*}
\begin{center}
 \includegraphics[clip,width=83mm,height=60mm]{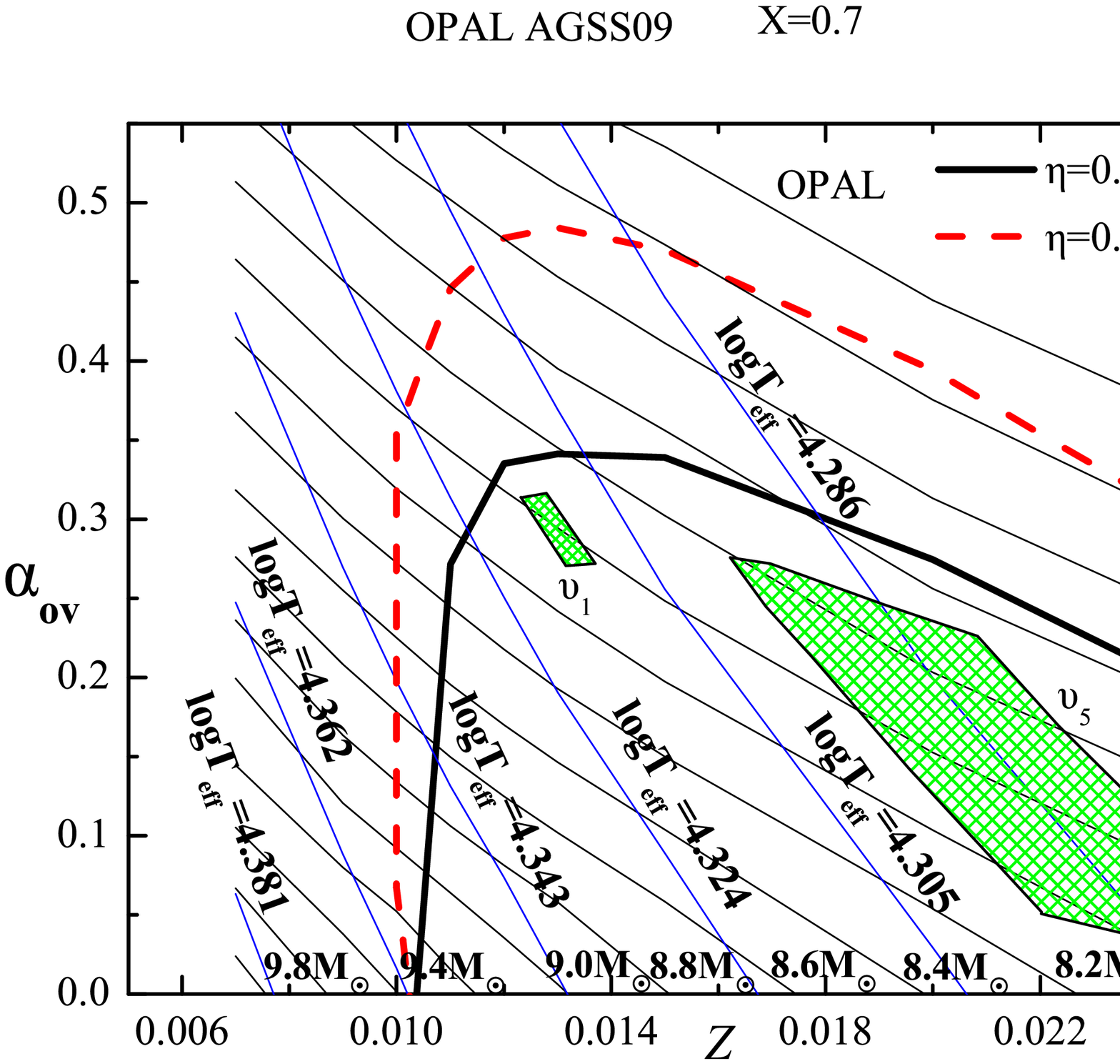}
 \includegraphics[clip,width=83mm,height=60mm]{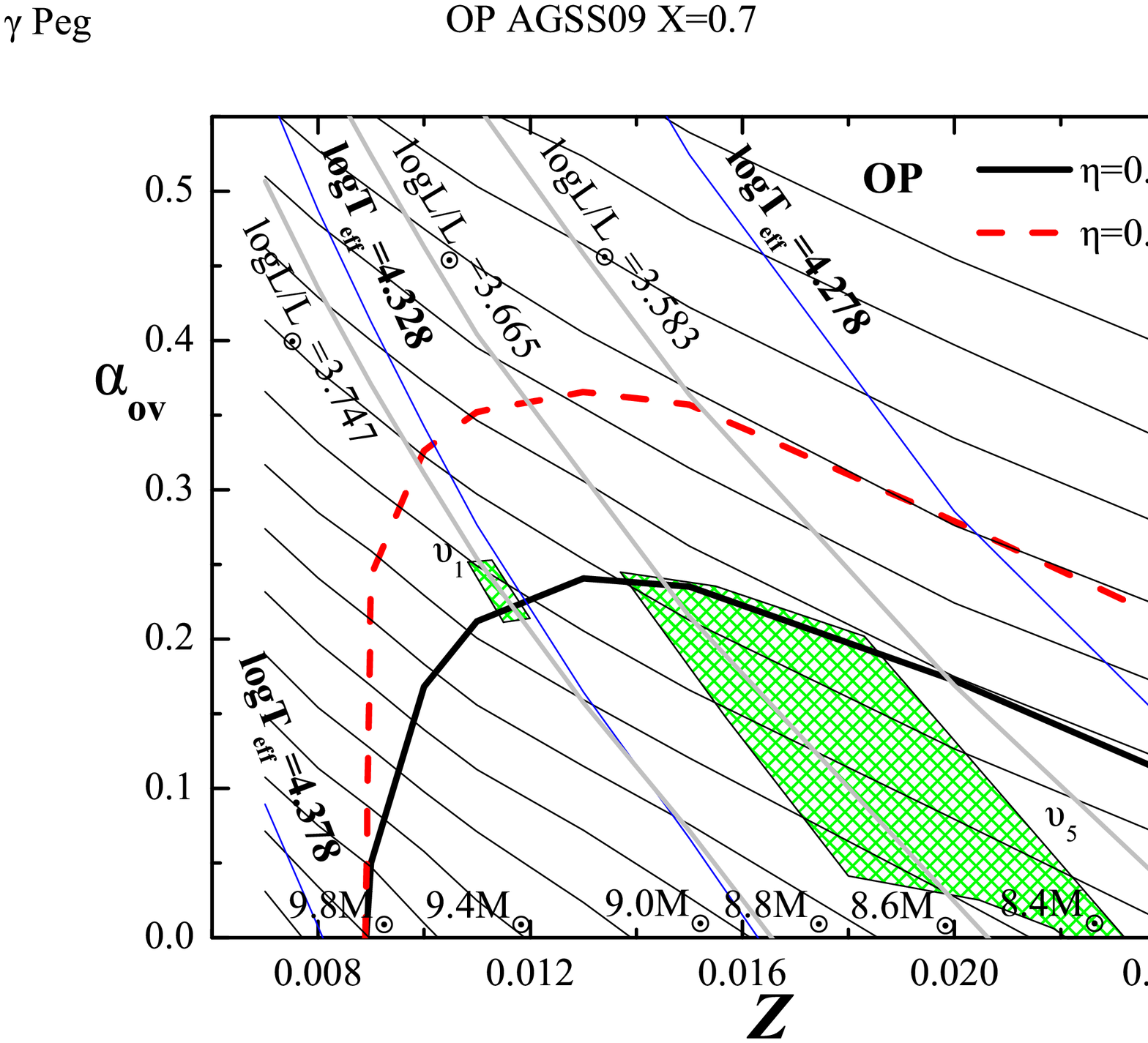}
 \caption{The overshooting parameter, $\alpha_{\rm ov}$, as a function of metallicity, $Z$, for the seismic models of $\gamma$ Peg found from the fitting of the $\nu_{1}$ frequency (the $\ell = 0$, $p_1$ mode) and the $\nu_5$ frequency (the $\ell = 1$, $g_1$ mode) for hydrogen abundance $X=0.7$, the OPAL (left panel) and OP opacities (right panel). Hatched areas indicates models fitting $f$-parameter
 for $\nu_1$ and $\nu_5$.}
\label{Z-AO-OPAL}
\end{center}
\end{figure*}
We depicted the instability borders with the thick solid line for the radial mode, $\nu_1$,
and with the thick dashed line for the dipole mode, $\nu_5$. Models that are located below the $\eta=0$ lines are unstable.
We plotted also the lines of constant masses and effective temperatures.
Results for the OP tables \citep{OP} are quite similar, as can be seen in the right panel of Fig.\,3.
In both panels, we marked also models which reproduce, within the observational errors,
the empirical values of $f$ (hatched areas) of the radial fundamental mode (labeled as $\nu_1$)
and of the dipole mode g$_1$ (labeled as $\nu_5$). The region covering models fitting the $f$-parameter
corresponding to $\nu_5$ is quite large because of larger observational errors in amplitudes and phases.
Unfortunately, there is no model fitting the $f$-parameter for $\nu_1$ and $\nu_5$ simultaneously,
regardless of which opacity tables are used.
Changing hydrogen abundance did not help in achieving an agreement.

\begin{figure}
\begin{center}
\includegraphics[clip,width=83mm,height=60mm]{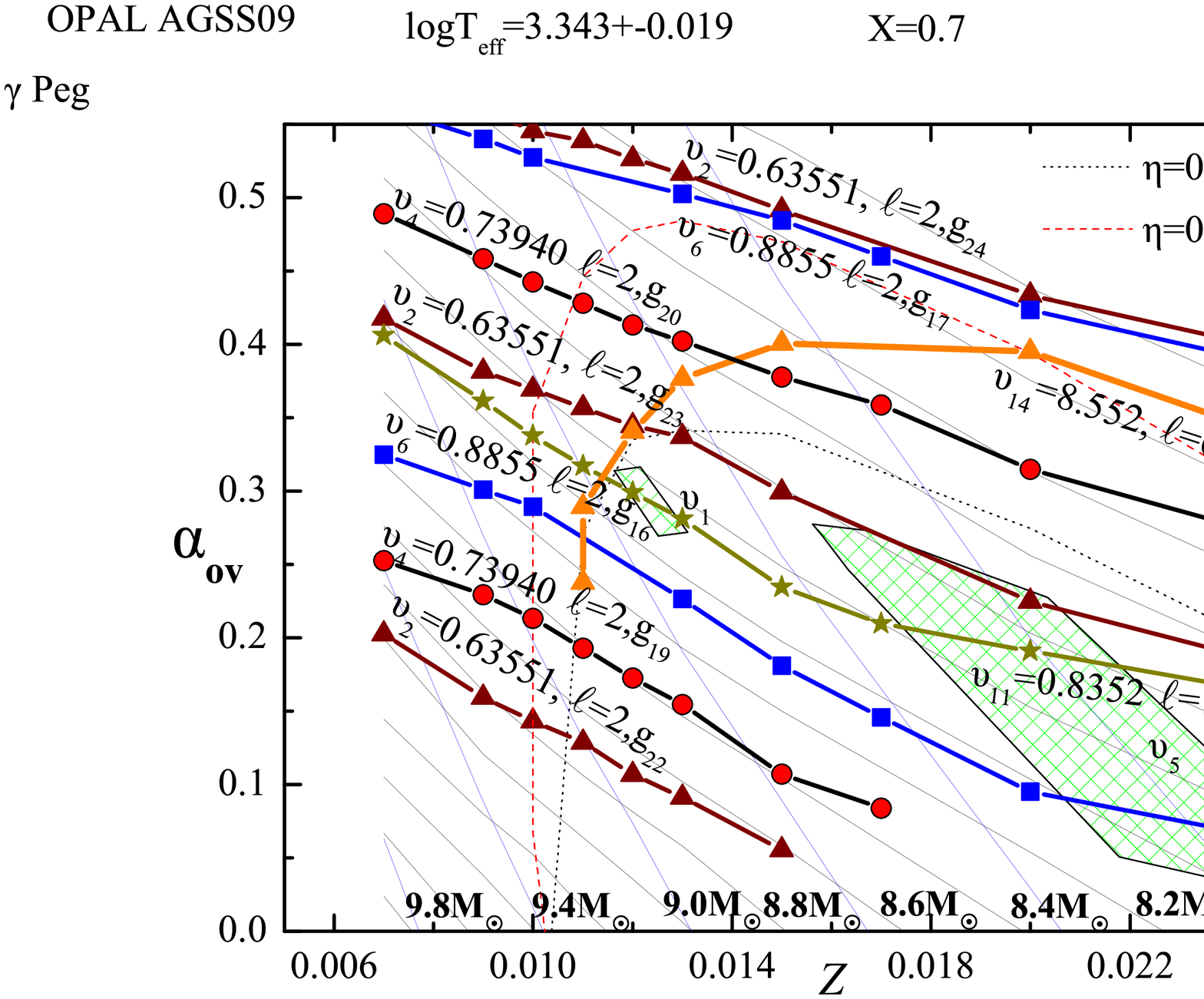}
\caption{The same as in the left panel of Fig.\,\ref{Z-AO-OPAL} but we marked seismic models fitting also
the high-order g-mode frequencies $\nu_2$, $\nu_4$, $\nu_6$, $\nu_{11}$ and the high degree mode $\nu_{14}$.}
\label{Z-AO-OPAL-g}
\end{center}
\end{figure}

Let us now consider other well identified modes: $\nu_2$, $\nu_4$, $\nu_6$, $\nu_{11}$ and $\nu_{14}$.
If we assume that these modes are axisymmetric ($m=0$), then we can search for pulsational models reproducing
additionally one of these five frequencies.
In the case of high-order g-modes ($\nu_2$, $\nu_4$, $\nu_6$, $\nu_{11}$), we had to include more than one radial order.
In the considered range of the metallicity and overshooting parameter, the $\nu_2$ frequency
is $\ell=2$, g$_{22}$, g$_{23}$ or g$_{24}$ mode. The $\nu_3$ frequency is the $\ell=2$, g$_{19}$ or g$_{20}$ mode
and $\nu_6$ can be the $\ell=2$, g$_{16}$ or g$_{17}$ mode. We depicted also the $\ell=1$, g$_9$ mode corresponding to $\nu_{11}$,
and the $\ell=6$, g$_2$ mode corresponding to $\nu_{14}$. Results are presented in Fig.\,\ref{Z-AO-OPAL-g}.
As we can see, models reproducing $\nu_{11}$ cross the hatched areas described above.
This means that there are models fitting three frequencies: $\nu_1$, $\nu_5$, $\nu_{11}$ and the $f$-parameter
for the radial fundamental or dipole g$_1$ mode. Lines of the $\nu_2$ and $\nu_6$ modes run through the hatched region of $\nu_5$.

Although we managed to derive the empirical values of the $f$-parame\-ter for the high-order g-modes, we could not find models reproducing simultaneously the real and imaginary part of $f$ for any of the SPB-type mode, neither with the OPAL nor OP opacities.

\section{Conclusions}

The aim of this paper was to construct seismic models of $\gamma$ Peg which reproduce both pulsational frequencies
and corresponding values of the nonadiabatic complex parameter $f$. Although we did not fully succeed,
we have showed directions and problems that need to be solved.

The first problem we encountered, was that there was no seismic model reproducing simultaneously
the $f$-parameter of the radial ($\nu_1$) and dipole ($\nu_5$) modes.
Secondly, there was no seismic model fitting empirical values of the $f$-parameter
for any high-order g-mode.

These inconsistencies can be caused either by the underestimated errors or/and indicate
that some additional effects should be included in pulsation modeling.
One obvious solution could be inadequacies in the opacity data.
Recently, \citet{ZP2009} have suggested increasing opacity by 20-50\%
around the Z-bump and DOB (Deep Opacity Bump) to explain the observed frequency range of $\gamma$ Peg.
Our further studies will show whether this can help also to solve problems reported in this paper.

\acknowledgements
We gratefully thank Gerald Handler for \linebreak kindly providing data
on photometric and radial velocity variations.
This work was supported by the HELAS EU Network, FP6, No. 026138
and the Polish MNiSW grant N N203 379636.

\end{document}